# Mathematical link of evolving aging and complexity

Byung Mook Weon[1,2][*] and Jung Ho Je[2][*]

[1] *Department of Physics, School of Engineering and Applied Sciences, Harvard University, 9 Oxford Street, Cambridge, MA 02138, USA.*

[2] *Department of Materials Science and Engineering, X-ray Imaging Center, Pohang University of Science and Technology, Pohang, 790-784, Korea.*

**Corresponding Author**

Dr. Byung Mook Weon : bmweon@hotmail.com
Prof. Jung Ho Je : jhje@postech.ac.kr

**ABSTRACT**

**Aging is a fundamental aspect of living systems that undergo a progressive deterioration of physiological function with age and an increase of vulnerability to disease and death. Living systems, known as complex systems, require complexity in interactions among molecules, cells, organs, and individuals or regulatory mechanisms to perform a variety of activities for survival. On this basis, aging can be understood in terms of a progressive loss of complexity with age; this suggests that complexity in living systems would evolve with age. In general, aging dynamics is mathematically depicted by a survival function, which monotonically changes from 1 to 0 with age. It would be then useful to find an adequate survival function to link aging dynamics and complexity evolution. Here we describe a flexible survival function, which is derived from the stretched exponential function by adopting an age-dependent exponent. We note that the exponent is associated with evolving complexity, i.e., a fractal-like scaling in cumulative mortality. The survival function well depicts a general feature in survival curves; healthy populations show a tendency to evolve towards rectangular-like survival curves, as examples in humans or laboratory animals. This tendency suggests that both aging and complexity would evolve towards healthy survival in living systems. Our function to link aging with complexity may contribute to better understanding of biological aging in terms of complexity evolution.**

# Introduction

Living systems are known as open *self-organizing* complex adaptive systems [1] and must often cope with hostile environmental conditions for *survival* [2]. To maintain their *far-from-equilibrium* (living) state, living things should exchange matter, energy, and information from their surroundings and further adapt themselves to genetic and environmental fluctuations [1, 2]. To perform a variety of activities for survival, living systems require complexity in interactions among molecules, cells, organs, and individuals or regulatory mechanisms [3]. A hallmark of living systems is their extraordinary *complexity* in physiological function [3]. The concept of complexity, derived from the field of *nonlinear dynamics* in physics and mathematics, is useful to measure the output of complex physiologic processes in biology and medicine [3]. *Healthy* living systems with complexity are characterized by *adaptive* interactions of multiple control mechanisms that enable each individual to adapt to the exigencies and unpredictable changes of everyday life [3]. Generally, *aging* can be understood as a progressive loss of complexity with age in physiological systems [3]. The formulation of *survival* or *mortality curves* is essential for the quantification of aging dynamics to all scientists who study aging, such as demographers, biologists, and gerontologists. A single survival curve reflects a variance in survival probability (equally, survival function and survival rate) as a function of age. Despite recent physical approaches to population biology [4–6], a general picture of aging dynamics associated with complexity evolution of living systems remains elusive.

Many mathematical models for survival curves have been proposed (see recent reviews in refs. [7, 8]). One of the fundamental mortality laws is the Gompertz law [9], in which the mortality rate increases roughly exponentially with increasing age at senescence. However, it seems to be obvious that the human mortality rate does not increase according to the Gompertz law at very old ages [8, 10] and the deviation from the Gompertz law remains a great puzzle to demographers, biologists, and gerontologists. Many other mathematical models such as the Weibull, the Heligman-Pollard, the Kannisto, the quadratic, and the logistic models yet provide poor fit to the empirical mortality patterns at very old ages [8]. There are still needs for an appropriate mathematics for survival (or mortality) curves with simplicity, efficiency, and flexibility in complexity analysis [11–14]. In previous works [15–17], we put forward a universal survival function, which is derived from the stretched exponential function [18–21]. In this study, we address that this function enables us to connect aging dynamics to complexity evolution.

The stretched exponential function is widely used to describe complex dynamics in physics and biology [18–21]. In physics, *relaxation* is an aging process in which a system gradually changes from a far-from-equilibrium (living) to an equilibrium (dead) state. Structural relaxation of a glassy state towards a metastable equilibrium amorphous state is often referred as "physical aging", which generally exhibits *nonexponential* relaxation [20, 21]. The temporal behavior of the response function $s(u)$ can be described by the *stretched exponential* or the *Kohlrausch-Williams-Watts* (KWW) function [22–24] (sometimes called the *Weibull* function [25]). This function has a general form of $s(u) = \exp(-u^{\beta})$ ($\beta > 0$) where $s(u)$ is the measurable quantity decreasing with age $u$ ($= x/\alpha$) where real age $x$ can be rescaled with a characteristic life $\alpha$ taken at $s(α) = \exp(-1) ≈ 0.3679$ [15]. The KWW function appears in many complex systems from soft matter systems, such as glass-forming liquids and amorphous solids [26, 27], to astrophysical objects [28]. The KWW function is typically classified as the "stretched" exponential for $0 < β < 1$, the "compressed" exponential for $β > 1$, and the "simple" exponential for $β = 1$. The nonexponential nature ($β ≠ 1$) is known to be related to the "dynamic heterogeneity" or the "fractal time" of the system [29]. In biology, the typical survival curves, $s(u)$, for humans and animals fall into three main types, usually known as Type-I, -II, and -III curves [30]. Type-I survival curves slightly change at early and middle ages and then suddenly decline at late ages, as seen for long-lived humans. Type-II curves almost linearly decease with age, as seen for short-lived birds. Type-III curves quickly decrease at early ages, as seen for most plants. Interestingly, Type-I survival curves resemble the compressed exponential curves ($β > 1$). Type-II curves are similar to the standard exponential curves ($β = 1$). Type-III curves correspond to the stretched exponential curves ($0 < β < 1$). In this study, we describe a new survival function, which is well adaptable to biological survival curves by modifying the classic KWW function.

## Results

### 1) Modeling Survivability

In our previous works, we showed that complicated survival curves are well described by modifying the KWW function with an "age-dependent" *shaping exponent* $β(u)$, as follows [15–17]:

$$s(u) = \exp(-u^{β(u)})$$

In this work, we argue that the age dependence of $β(u)$ can be understood in terms of *fractal time* $u^{β(u)}$ replacing time $u$ [29]. Conceptually, fractal time describes highly intermittent self-similar temporal behavior that does not possess a characteristic time scale [29, 31]. The cumulative mortality (hazard) function, $h(u) = -\ln(s) = u^{β(u)}$, (defined as the negative logarithmic survival function, equally the fractal time in our case) shows a typical *fractal-like scaling* as $h(bu) = b^{β(u)}h(u)$ for a constant $b$. By rescaling $u$, instead of $x$, we are able to discover a universal, species-independent, scale-invariant survival law from different species. Interestingly, we note that $β(u)$ can vary with time in physical systems, for instance, in luminescence decays [15] or carrier hopping dynamics [32]. Most importantly, we emphasize that the shaping exponent $β(u)$ is a dynamic *variable* (i.e., *dynamically evolves*) with age for biological systems. The age-related evolution of $β(u)$ would shed light on how living systems *evolve* towards their "slowest aging rates" for healthy survival, which would be a key feature of evolving aging dynamics in living systems.

We describe the basic assumptions of our mathematical approach, as illustrated in **Fig. 1**. We assume three points as follows. (1) Living systems continually try to survive by adapting themselves to internal or external factors, including genetic (biological) and non-genetic (environmental) changes. (2) Complexity in living systems evolves for better survival towards healthy aging. Complexity evolution would arise from all functional interactions, as shown by the arrows in the left panel, among genes, molecules, cells, organs, and individuals within living systems or with environments. Consequently, (3) a survival function depicts aging dynamics and complexity evolution of an individual or a population evolving towards healthy aging. In fact, historic trends in human survival curves show a gradual evolution toward healthy aging. We are able to describe those trends in human survival curves with the stretched exponential function $s(u) = \exp(-u^{\beta(u)})$ by adjusting the age-dependent $\beta(u)$. Biological differences from species (scaling effect) may be cancelled by rescaling age (the $x$ axis) as illustrated in the right panel of **Fig. 1**. Using the age-dependent, shaping exponent, $\beta(u)$, we show that survival curves evolve towards the rectangular-like curves, defined as the ideal healthy aging. Here we believe that $\beta(u)$ has mathematical clues to link evolving aging and complexity for living systems.

## 2) Slowest Aging Rates

We think of the concept of "healthy aging", which may be indeed an ultimately adaptive feature in survival curves of living systems. As a typical example, we consider survival curves for humans. From life table data of human populations in developed countries, we find that survival probability evolves over time (remarkably, for last several decades), universally towards a *rectangular* shape, as many scientists observed [33–35]. Such a gradual evolution in human survival curves, as called the "rectangularization" [33], results from the fact that survival curves become more rectangular and deaths are compressed to higher ages. How can we describe the evolving aging of living systems? Presumably the healthy aging can be characterized as the "slowest aging rates", equivalently indicating that the mortality rates are minimized at each age. Here we suggest a simple mathematical approach, which would be useful to test the validity of the rectangularization hypothesis.

By modeling $s(u)$ as a modified stretched exponential function with an age-dependent shaping exponent, we are able to evaluate the slowest aging rates [16, 17]. Generally speaking, survival function $s(u)$ decreases monotonically from 1 to 0 with age; that is, $-ds(u)/du > 0$. Let $-ds(u)/du \to 0$ denote the slowest aging rates, equivalently implying the largest survival rates ($\approx 1$ at $u < 1$). We find that $-ds(u)/du \to 0$ can be achieved only for a positive slope of $\beta(u)$ ($d\beta/du > \varepsilon$) early in life ($u < 1$) or for a negative slope of $\beta(u)$ ($d\beta/du > \varepsilon$) at late life ($u > 1$), since the quantity $\varepsilon = -\beta/u\ln(u)$ is positive at $u < 1$ and negative at $u > 1$ [16, 17]. Using the age dependence of $\beta(u)$, we are able to describe the slowest aging rates in living systems.

Here we derive a simple mathematics to determine the slowest aging rates. By solving the equation $d\beta/du = -\beta/u\ln(u)$ to satisfy "$-ds(u)/du = 0$", we obtain an ideal curve of $\beta(u)$ that holds if and only if the derivative of the survival probability approaches zero, implying the "slowest aging rates". This condition reflects the rectangular survival curve, where $s(u)$ becomes completely rectangular in which all individuals are alive with $s(u) \approx 1$ to its characteristic life ($u = 1$) and suddenly die with $s(u) \approx 0$ after $u > 1$. Therefore the slowest aging rates of living systems can be simply described as:

$$\beta(u) = \frac{c}{|\ln(u)|}$$

Here, $c$ is a universal constant (numerically $c \geq 7$ for $s > 0.99908$ at $u < 1$). Most importantly, the rectangular survival curve can be precisely approximated with $\beta(u)$ for $c = 7$. This gives a criterion that whenever living systems evolve to achieve the maximum survival probability, $\beta(u)$ should evolve towards the ideal curve, $\beta(u) = 7|\ln(u)|^{-1}$. Therefore, $\beta(u)$ is a good measure to determine the appearance of the rectangularity in survival curves.

### 3) Application and Validation

We test the validity of our mathematical approach in real systems. For typical human survival curves, the age-related trajectory of $\beta(u)$ increases with age towards the ideal curve of $\beta(u)$ at early ages ($u < 1$), as fitted by a linear form of $\beta_0 + \beta_1 u$, and in turn decreases with age at late ages ($u > 1$), as fitted by a quadratic form of $\beta_0 + \beta_1 u + \beta_2 u^2$ [16, 17]. For instance, from Swedish female periodic life table in 2000, we roughly estimate $\beta(u)$ with $\beta_0 \approx 1$ and $\beta_1 \approx 9$ at $u < 1$ and $\beta_0 \approx -20$, $\beta_1 \approx 60$, and $\beta_2 \approx -30$ at $u > 1$, as depicted as circles in **Fig. 2**. As illustrated in the inset, the survival curve obtained from the estimated $\beta(u)$ (circles) quite well agrees with the actual survival curve of Swedish females in 2000 (dashed line). Remarkably, the estimated and the measured survival curves are very close to the rectangular-like survival curve (solid line).

  The evolution towards the slowest aging rates makes survival curves to have rectangularity. We test this feature for humans with the reliable demographic data taken from the Human Mortality Database (http://www.mortality.org). The survival curves are analyzed by computing the age-dependent shaping exponents $\beta(u) = \ln(-\ln(s))/\ln(u)$ with rescaled age ($u = x/\alpha$) from the general expression of survival probability $s = \exp(-u^{\beta(u)})$, where the characteristic life ($\alpha$) is graphically taken at $s(\alpha) = \exp(-1)$ for each survival curve. We find that Swedish females' survival curves during last two centuries from 1800 to 2000 corroborate that the age-related trajectory of $\beta(u)$ evolves in a fashion predicted in theory (i.e., towards the ideal curve, $\beta(u) = 7|\ln(u)|^{-1}$, solid line), as illustrated in **Fig. 3**. The age-dependent evolution of $\beta(u)$ provides conclusive evidence that human survival curves evolve towards the slowest aging rates, which supports the validity of the rectangularization hypothesis [33].

  Let us consider mortality rate (sometimes called hazard function or force of mortality), defined as a negative derivative of logarithmic survival rate). The age-related evolution of $\beta(u)$ has clues to interpret the age-related evolution of mortality rate. The mortality rate $\mu(u) = -d\ln(s)/du$ in our model is described with respect to the age-dependent $\beta(u)$ as:

$$\mu(u) = \left(\frac{u^{\beta(u)}}{\alpha}\right)\left[\frac{\beta(u)}{u} + \ln(u)\frac{d\beta(u)}{du}\right]$$

Here, the mortality rate can be calculated from $\alpha$ and $\beta(u)$ for a single survival curve. For ideal survival conditions, $\beta(u)$ tends to evolve towards the ideal curve of $\beta(u)$ and the mortality rate should become zero for all ages. For real survival conditions (for humans or animals), as the survival curve approaches to a rectangular shape, $\beta(u)$ tends to linearly increase with age at early ages ($u < 1$), which corresponds to a linear increase of $\ln(\mu)$. The linearity in $\ln(\mu)$ at middle ages ($u = 0.5$–$1.0$, equivalently 30–80 years for humans) is similar to the *Gompertz* mortality law [17]. At late ages ($u > 1$), $\beta(u)$ shows a quadratic change with age, which corresponds to a quadratic change of $\ln(\mu)$. The quadratic decline of $\ln(\mu)$ is practically valid for the oldest-age mortality patterns [34]. We find a general scaling relation between $\beta(u)$ and $\ln(\mu)$; that is, $\beta(u)$ is approximately proportional to $\ln(\mu)$, which is valid as shown in **Figs. 3 and 4**.

In principle, the mortality rate is predicted to reach zero for all ages, if the survival rate approaches to the rectangular curve. This feature can be identified by evaluating the age-related evolution of $\beta(u)$, which gets close to the "ideal" curve of $\beta(u) = 7|\ln(u)|^{-1}$. In practice, it would be relatively easy to diminish the mortality rate to a very low level at young ages through health care and welfare but not at old ages by age-associated progressive aging. We find that the mortality rates of the oldest seem to be fixed at $\sim 1$ (parallel shift, ←) over time for humans, while the infant mortality rates plastically fall downward (vertical shift, ↓), as shown in **Fig. 4**. Such parallel and vertical shifts of the mortality rates imply that death is being delayed because people are reaching old age in better health [36]. We also find a convergence of $\mu(u)$ at old ages as depicted in **Fig. 4**. This convergence would universally exist in living systems [37] if death rates become relatively similar at old ages for a single species.

We further investigate the age-related evolution of $\beta(u)$ for non-human systems (**Fig. 5**). The reliable life tables were taken for wild-type flies (*Drosophila melanogaster*, $\alpha = 46.38$ days) [38], worms (*Blatta orientalis*, $\alpha = 45.39$ days) [38], automobiles ($\alpha = 8.08$ yrs) [38], and tyrannosaurs (*Albertosaurus sarcophagus*, $\alpha = 17.45$ yrs) [39]. The data were arbitrary selected for analysis. Indeed, $\beta(u)$ dynamically evolves with age for living systems, which is remarkably shown in flies, while $\beta(u)$ is relatively constant ($\beta \approx 2$) for automobiles as non-living systems. Most importantly, the age-related trajectory of $\beta(u)$ for flies is very similar to that for humans, approaching to the ideal curve, $\beta(u) = 7|\ln(u)|^{-1}$. Additionally, the mortality rate of flies at $u = 0.5$–$1.0$ linearly increases with age, showing the validity of the Gompertz law, and also comes to a saturation at $u > 1$, showing the singularity of the mortality plateau. These results suggest that the common features of mortality curves such as the Gompertz law and the mortality plateau might spontaneously emerge from the age-dependent shaping exponents that dynamically evolve towards the slowest aging rates of living systems.

For both cases of humans and flies, $\beta(u)$ gradually evolves with age, approaching to the ideal curve, $\beta(u) = 7|\ln(u)|^{-1}$, as predicted in theory. The universality of mortality patterns is well confirmed for insects, worms, and yeasts, as well as humans in many studies (e.g., see refs. [5, 34]). It is important to note that the evolution towards the maximum survival probability is *scale-invariant* regardless of biological species. This suggests that the evolution towards the slowest aging rates would be a species-independent, scale-invariant, universal aspect in survival dynamics of living systems.

**Discussion**

Why does the dynamic nature in shaping exponent *β*(*u*) emerge in living systems? Living systems are ubiquitous in our daily life: specifically, if processes underlying *life* are explained as a tendency towards *maximum entropy* production, systems such as galaxies and hurricanes might be described as living systems [40]. Living systems are capable of altering strategies and of adapting to internal and external changes. Healthy physiological functions require an integration of complex networks of control systems and feedback loops that operate on multiple scales in space and time [3]. These fractal-like physiological processes enable an organism to adapt to the exigencies of everyday life. In general, the complexity of living processes should be evolvable with environmental conditions. Adaptive interactions of each individual occur in a cooperative manner with its neighbors, resulting in dynamic self-organization and influencing flexible survival strategies between cooperation and competition. Since living systems would evolve to optimize their capabilities and strategies for survival, the slowest aging rates would spontaneously emerge. Presumably, living processes would not follow simple physical relaxation processes, where the shaping exponent is invariant with age. To illustrate evolvable survival curves of living systems, we adopt an age-related shaping exponent, *β*(*u*), which provides a simple criterion for the slowest aging rates. On this basis, we argue that dynamic change in shaping exponent is attributed to "complexity evolution" in living systems. On the other hand, complexity evolution would be able to give hints about how and why vitality declines with age [41]. Aging can be regarded as a loss of complexity in a variety of fractal-like anatomic structures and physiological processes, leading to a decline in adaptive capacity and ultimately a development of frailty at old ages [3]. Consequently, the appearance of the age-dependent *β*(*u*) in living systems is due to the evolution towards the slowest aging rates and fundamentally attributed to the complexity evolution (age-related evolution in complexity) for living systems.

**Conclusion**

In summary, we put forward a simple mathematics to link complexity evolution with aging dynamics of living systems. Our mathematics is useful to illustrate how living systems evolve towards the rectangular survival curves and the slowest aging rates. Such an elegant evolution is testable with a simple mathematical criterion derived from the age dependence of the shaping exponent in the stretched exponential survival function. The feature of the slowest aging rates emerges in survival curves for humans and flies. The age-dependent shaping exponent is a good indicator of complexity evolution (age-related evolution in complexity), which is associated with evolving fractal-like scaling in cumulative mortality rates. Our finding may give useful insights into basic aspects of aging in living systems by offering a mathematical linkage between aging dynamics to complexity evolution.

# Figures

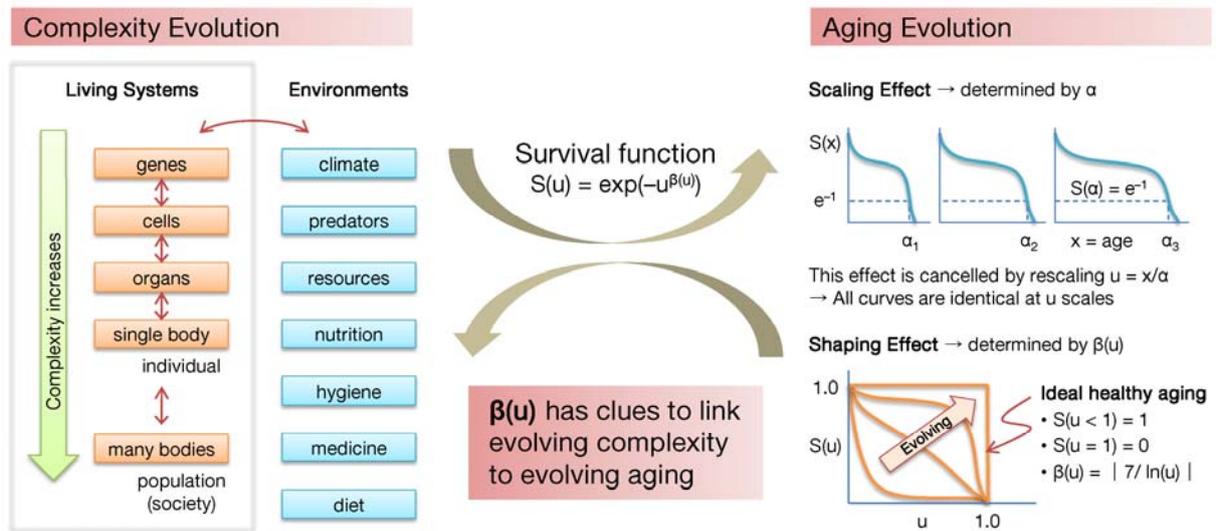

**Figure 1: Basic assumptions.** Left panel: (1) Living systems continually try to survive by adapting themselves to internal (genetic) or external (non-genetic) changes. (2) Complexity in living systems evolves for better survival towards healthy aging, through all functional interactions among genes, molecules, cells, organs, and individuals within living systems or with environments. Right panel: (3) Survival curves reveal aging dynamics and complexity evolution in an individual or a population evolving towards healthy aging. Biological differences from species (scaling effect) can be cancelled by rescaling age (the $x$ axis). Middle panel: Survival curves are well described with the stretched exponential survival function $s(u) = \exp(-u^{\beta(u)})$ (see the details in the text) by adjusting the age-dependent, shaping exponent, $\beta(u)$. Here $\beta(u)$ has clues to link evolving complexity and evolving aging for living systems.

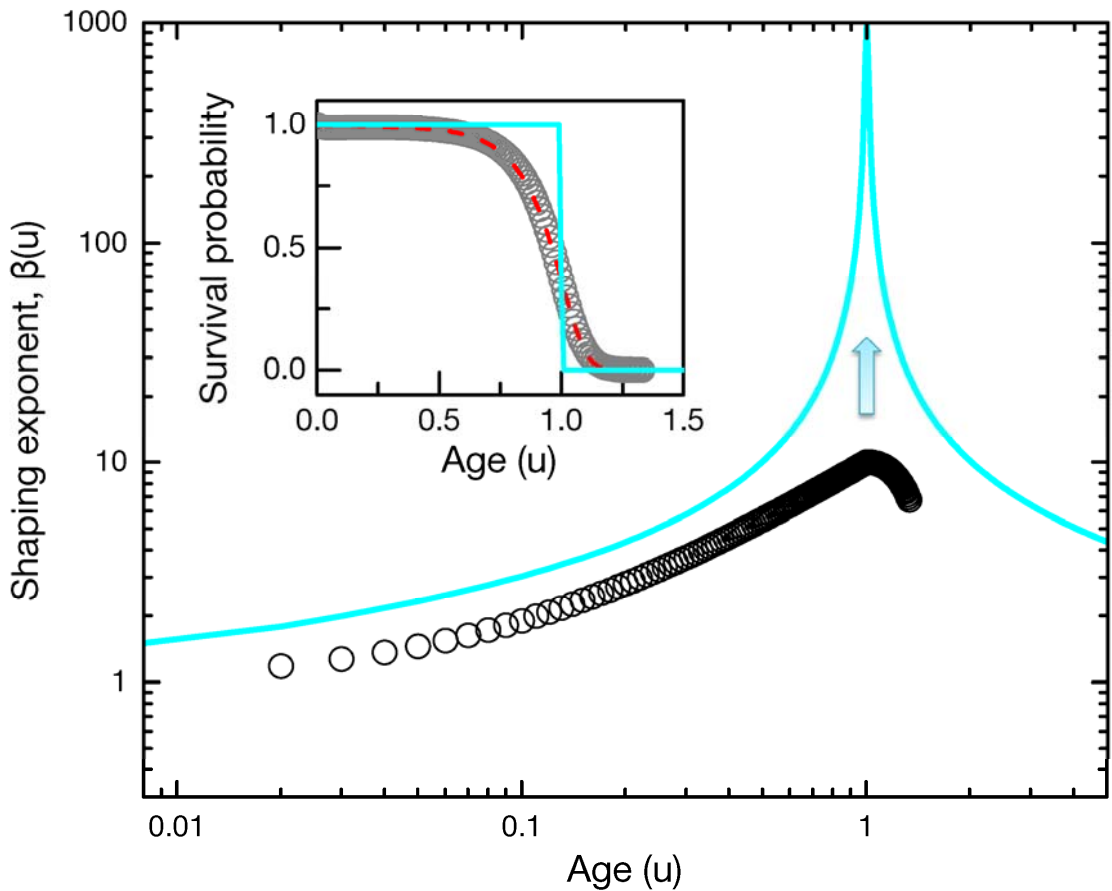

**Figure 2: Healthy survival curves.** The shaping exponent, *β*(*u*), approaches to the ideal curve, $\beta(u) = 7|\ln(u)|^{-1}$ (solid line), as the survival probability, *s*(*u*), approaches to the rectangular shape (solid line, inset). This tendency, which stands for healthy aging, is obvious in a typical trajectory of *β*(*u*) for humans (circles, see the fitting parameters in the text). The survival curve estimated from the typical *β*(*u*) (circles, inset) agrees well with the real survival curve for Swedish females in 2000 (dashed line, inset).

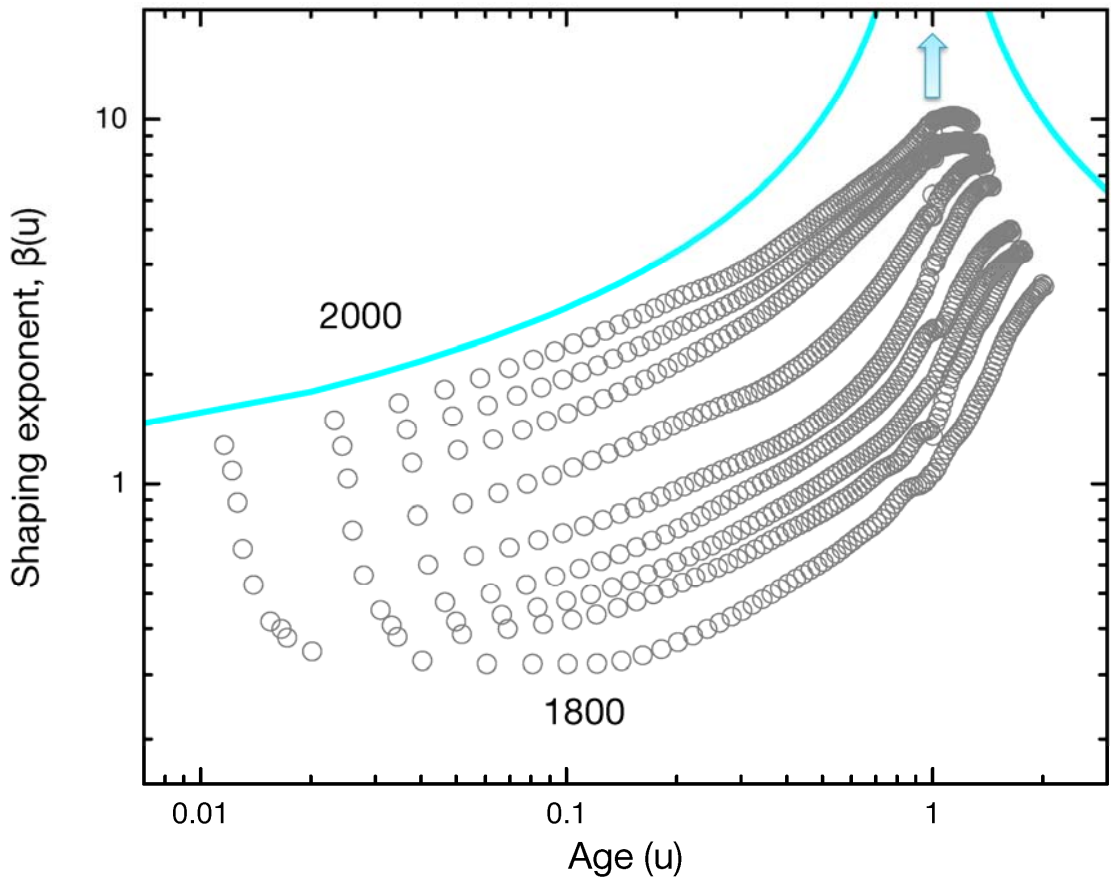

**Figure 3: Evolution of shaping exponents for humans.** The historic evolution in survival curves for Swedish females shows that $\beta(u)$ dynamically evolves towards the slowest aging rates ($\beta(u) = 7|\ln(u)|^{-1}$, solid line), where $s(u)$ becomes rectangular for last two centuries (1800–2000 years).

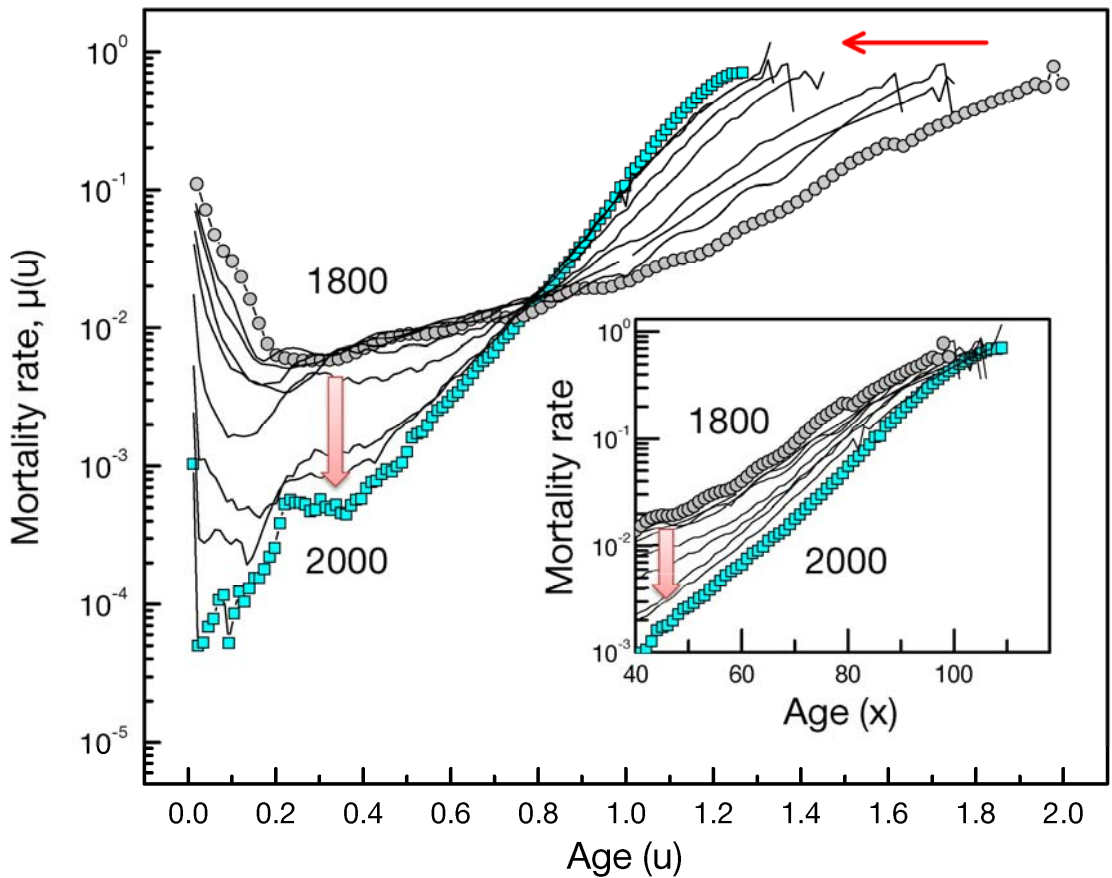

**Figure 4: Evolution of mortality rates for humans.** The mortality curves for Swedish females for the last two centuries demonstrate that human mortality curves evolve towards the slowest aging rates. Interestingly, at late ages the mortality seems to be fixed (parallel shift, ←) over time, while at early ages it significantly falls down (vertical shift, ↓). Here, $\mu(u)$ linearly increases with reduced age ($u$) at middle ages, showing the Gompertz law, and converges with real age ($x = \alpha u$ [years] with its characteristic life $\alpha$) at old ages (inset).

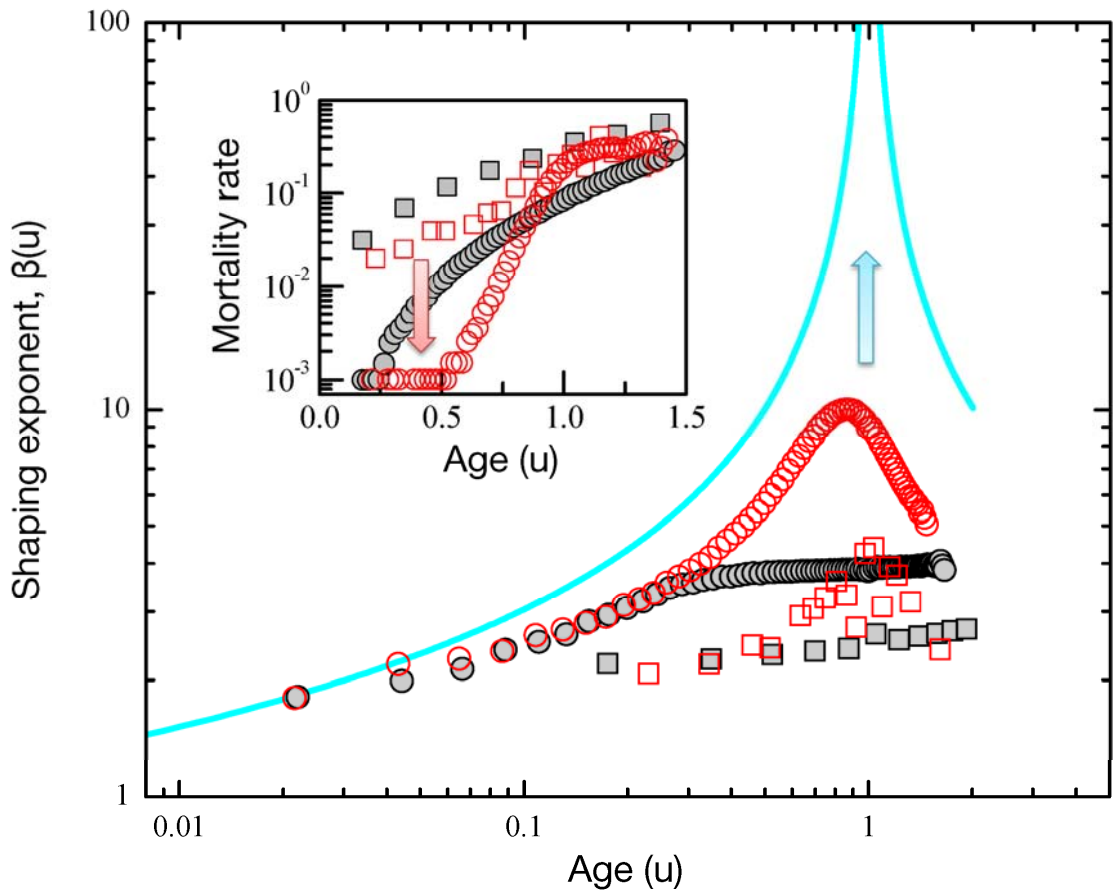

**Figure 5: Evolutions of shaping exponents and mortality rates for non-human systems.** The feature of the slowest aging rates (solid line, $\beta(u) = 7|\ln(u)|^{-1}$) emerges for wild-type flies (*Drosophila melanogaster*, open circles), worms (*Blatta orientalis*, closed circles), and tyrannosaurs (*Albertosaurus sarcophagus*, open squares). In contrast, automobiles (closed squares) show a constant value of $\beta$ and a high value of $\mu$, indicating typical non-living systems.